\documentclass[fleqn,11pt]{wlscirep}
\usepackage[utf8]{inputenc}
\usepackage[T1]{fontenc}
\usepackage{hyperref}
\usepackage{authblk}
\usepackage{xcolor}
\usepackage{ragged2e}
\usepackage{amsmath}
\usepackage{siunitx}
\usepackage{hyperref}
\newcommand{\bm}{\mathbf}
\usepackage[version=4]{mhchem}
\newcommand{\R}{\mathbb{R}}

\definecolor{myblue}{rgb}{0,0,255}
\definecolor{myred}{rgb}{255,0,0}

\title{Enhancing Materials Discovery with Valence Constrained Design in Generative Modeling}

\author[1,2,3,$\dagger$,*]{Mouyang Cheng}
\author[4,$\dagger$]{Weiliang Luo}
\author[3,$\dagger$]{Hao Tang}
\author[5]{Bowen Yu}
\author[6]{Yongqiang Cheng}
\author[7]{Weiwei Xie}
\author[2,3,8]{Ju Li}
\author[2,4,9]{Heather J. Kulik}
\author[1,2,8,**]{Mingda Li}
\affil[1]{Quantum Measurement Group, MIT, Cambridge, MA 02139, USA}
\affil[2]{Center for Computational Science $\&$ Engineering, MIT, Cambridge, MA 02139, USA}
\affil[3]{Department of Materials Science and Engineering, MIT, Cambridge, MA 02139, USA}
\affil[4]{Department of Chemistry, MIT, Cambridge, MA 02139, USA}
\affil[5]{Department of Physics, MIT, Cambridge, MA 02139, USA}
\affil[6]{Neutron Scattering Division, Oak Ridge National Laboratory, Oak Ridge, TN 37831, USA}
\affil[7]{Department of Chemistry, Michigan State University, East Lansing, MI 48824, USA}
\affil[8]{Department of Nuclear Science and Engineering, MIT, Cambridge, MA 02139, USA}
\affil[9]{Department of Chemical Engineering, MIT, Cambridge, MA 02139, USA}

\affil[$\dagger$]{These authors contributed equally.}
\affil[*]{e-mail: vipandyc@mit.edu}
\affil[**]{e-mail: mingda@mit.edu}

\begin{abstract}
Diffusion-based deep generative models have emerged as powerful tools for inverse materials design. Yet, many existing approaches overlook essential chemical constraints such as oxidation state balance, which can lead to chemically invalid structures. Here we introduce CrysVCD (Crystal generator with Valence-Constrained Design), a modular framework that integrates chemical rules directly into the generative process. CrysVCD first employs a transformer-based elemental language model to generate valence-balanced compositions, followed by a diffusion model to generate crystal structures. The valence constraint enables orders-of-magnitude more efficient chemical valence checking, compared to pure data-driven approaches with post-screening. When fine-tuned on stability metrics, CrysVCD achieves 85\% thermodynamic stability and 68\% phonon stability. Moreover, CrysVCD supports conditional generation of functional materials, enabling discovery of candidates such as high thermal conductivity semiconductors and high-$\kappa$ dielectric compounds. Designed as a general-purpose plugin, CrysVCD can be integrated into diverse generative pipeline to promote chemical validity, offering a reliable, scientifically grounded path for materials discovery.

\end{abstract}
\begin{document}

\flushbottom
\maketitle

\thispagestyle{empty}

\section*{Introduction}
Materials design has long been a cornerstone of modern technology, driving advancements across diverse fields, from structural and chemical materials to energy and quantum technologies. 
The integration of computational methods has significantly enhanced this process, enabling the automated discovery of promising candidates \cite{olson1997computational,olson2000designing,oganov2019structure}, with recent advances in artificial intelligence (AI) accelerating these efforts further. 
In particular, inverse design has emerged as a transformative paradigm, facilitating the targeted generation of functional materials with specific properties while reducing the computational burden of exhaustive forward screening \cite{zunger2018inverse}. Among inverse design techniques, deep generative models have achieved remarkable success by capturing intricate structure-property relationships to enable conditional materials generation \cite{han2024ai,cheng2025ai}. 
Notable examples include CDVAE \cite{xie2021crystal}, DiffCSP \cite{jiao2023crystal}, and MatterGen \cite{zeni2025generative}, which demonstrate the capability of diffusion-based models to generate chemically diverse materials with novel crystal structures.
Another approach incorporates large language models (LLMs) in the materials design process, utilizing inverse design prompted by human natural language \cite{gruver2024fine,antunes2024crystal,yang2025generative}. All of these advanced deep generative architectures are opening a new era of automatic materials discovery.

Despite significant progress, current generative approaches still encounter challenges in practical solid state material discovery scenarios. One key issue is to ensure that generated materials adhere to fundamental chemical principles, such as oxidation states and stoichiometric balance. While these chemical rules can be learned from sufficiently large datasets, chemically implausible candidates can still be proposed by current state-of-the-art models.
Another issue is the thermodynamic stability of the generated materials. Methods such as guided generation \cite{ho2022classifier,dhariwal2021diffusion,karras2024guiding} and post-processing with density functional theory (DFT) help address this, although the DFT's computational cost, particularly for dynamical stability when evaluated with phonon calculations, limit the scalability. 
In addition, while models have shown success in generating materials with specific target properties such as bulk modulus and bandgap~\cite{zeni2025generative}, incorporating a wider range of functional characteristics, such as thermal and electrical conductivity, or optical behavior, could enhance their real-world utility.

In this work, we present CrysVCD (Crystal generator with Valence-Constrained Design), a deep generative architecture that combines diffusion and transformer models to generate \textit{ab initio} stable crystal structures while respecting fundamental chemical constraints.
Built upon recent advances such as MatterGen and DiffCSP, CrysVCD introduces a dedicated transformer module that explicitly ensures charge-balanced compositions. The generation process follows two steps: in Stage I, a transformer-based elemental language model outputs chemically valid formulas where each autoregressive token corresponds to a charged ion, explicitly enforcing oxidation state balance; in Stage II, a conditional diffusion model generates the corresponding atomic structure based on the formula. This design enables rapid screening of chemically plausible compositions during inference, ensuring that all generated crystals adhere to fundamental chemical valence constraints, an essential stability condition that has been underexplored. 
Within a comparable inference time, CrysVCD enables rapid screening of total chemical valence during Stage I, which guarantees balanced charge valence for all end product crystals.
The stability of generated crystals are evaluated on-the-fly using a pretrained machine learning interatomic potential (MLIP), in terms of both formation energy above the convex hull ($E_{\text{hull}}$) and phonon imaginary frequency.
By further tuning the model with labeled stability metrics, the conditional CrysVCD model achieves $>85\%$ stability rate of materials according to the criteria of $E_{\text{hull}}$ ($E_{\text{hull}}<0.1$ eV), and $68\%$ of materials that are stable as assessed by phonon calculations. 
We further apply CrysVCD to search for materials with high thermal conductivity and high-$\kappa$ dielectrics, both relevant for microelectronics applications.  
Overall, CrysVCD provides a general-purpose, physics-constrained framework that enhances the reliability and throughput of generative materials discovery. 

\section*{Results}
\subsection*{Model architecture}
We start by briefly reviewing recent diffusion-based generative modeling for materials design. A periodic crystal structure can generally be represented as $\bm{M} = (\bm{A}, \bm{X}, \bm{L})$, comprising atom types ($\textbf{A}$), fractional atomic coordinates ($\textbf{X}$), and lattice matrices ($\textbf{L}$).
Diffusion models progressively add Gaussian noise to these components over a sequence of time steps. A neural network, often instantiated as a graph neural network (GNN) representing the geometric structure \cite{xie2018crystal,reiser2022graph}, is trained to learn the reverse denoising process by approximating the score function \cite{ho2020denoising}. 
This learned score function enables the generation of diverse crystal structures by sampling from a noise distribution and iteratively denoising it back toward physically plausible configurations.
The current crystal structure generative models, such as DiffCSP and MatterGen, perform joint equivariant diffusion of $(\textbf{A}, \textbf{X}, \textbf{L})$. More specifically, the probability of generating $(\textbf{A}_0, \textbf{X}_0, \textbf{L}_0)$ can be expressed as
\begin{equation}
    p(\textbf{A}_0, \textbf{X}_0, \textbf{L}_0)=\prod_{t=1}^{T} \left(\int d\mathbf{M}_t\right)  p_T(\textbf{A}_{T}, \textbf{X}_{T}, \textbf{L}_{T})\prod_{t=T-1}^0  q_\theta(\textbf{A}_t, \textbf{X}_t, \textbf{L}_t|\textbf{A}_{t+1}, \textbf{X}_{t+1}, \textbf{L}_{t+1})
\end{equation}
where $\mathbf{M}_t=(\mathbf{A}_t, \mathbf{X}_t, \mathbf{L}_t)$ is the crystal structure at diffusion time step $t$, $q_\theta(\cdot)$ is a learnable function parametrized by an equivariant GNN, and $p_T(\cdot)$ is the prior noise distribution. 
This enables flexible structure generation and materials design.

\begin{figure}[!htbp]
    \centering
    \includegraphics[width=0.8\linewidth]{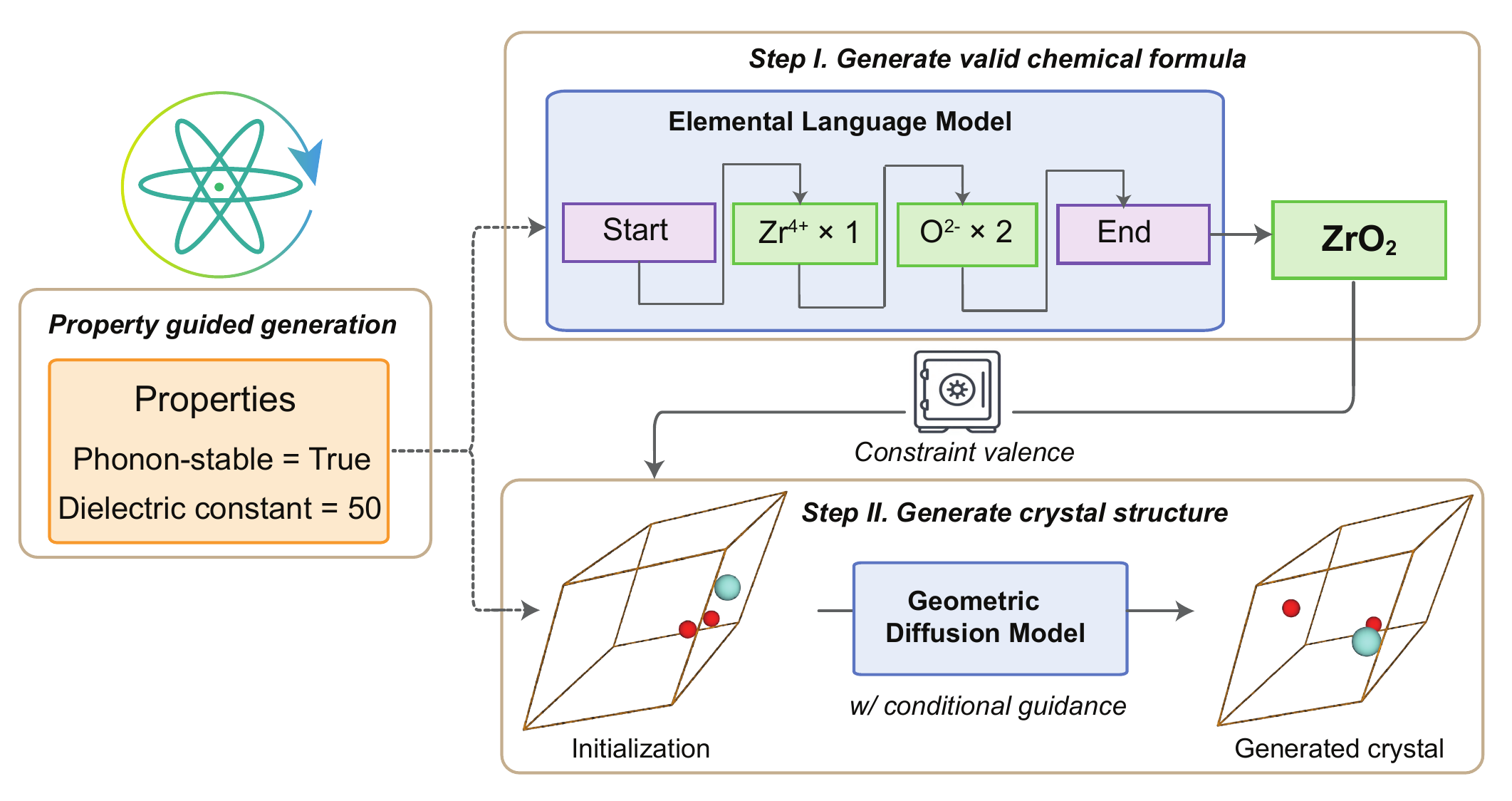}
    \caption{\textbf{Overview of CrysVCD}. Rather than performing pure data-driven denoising diffusion generation, CrysVCD introduces a two-step, property-guided pipeline that incorporates physical and chemical constraints. A property-conditioned elemental language model first generates a composition with balanced chemical valence (e.g., $\text{ZrO}_2$), which is then refined by a geometric diffusion model under classifier-free guidance, which interpolates conditional and unconditional outputs. Properties can encode thermodynamic stability as well as target functional performance.}
    \label{fig1}
\end{figure}

However, the above diffusion-based process is entirely data-driven and lacks explicit physical or chemical constraints. Current approaches typically first generate materials and down-select based on constraints, which reduces the efficiency. It would be desirable to impose fundamental constraints earlier in the process, e.g., before data generation. For instance, fundamental rules such as stoichiometry should apply to the atom types $\mathbf{A}$. This raises a key question: can we eliminate non-physical candidates before initiating the generative diffusion process? If so, this would enable a lightweight and scalable approach to incorporating domain-specific constraints early on. 
In light of this, we introduce CrysVCD, a two-step, property-aware crystal generation framework that explicitly incorporates physical and chemical constraints, as shown in Fig.\,\ref{fig1}.

CrysVCD decouples the generation of crystal compositions and geometries into two integrated and separately trained modules: I. an elemental language model guided by target properties and chemical valence rules, and II. a geometric diffusion model that generates 3D atomic structures consistent with the predicted composition and target properties.
More specifically, CrysVCD generates atomic species $\mathbf{A}_0$ first using the elemental language model, and then generates the corresponding crystal structure $\mathbf{X}$ and $\mathbf{L}$. Thus, the probability of generating $(\textbf{A}_0, \textbf{X}_0, \textbf{L}_0)$ for CrysVCD is expressed as
\begin{align}
    p(\textbf{A}_0, \textbf{X}_0, \textbf{L}_0)&= \prod_{t=1}^{T} \left(\int d\mathbf{X}_td\mathbf{L}_t\right) p(\textbf{A}_0)\cdot p_T(\textbf{A}_0, \textbf{X}_{T}, \textbf{L}_{T})\prod_{t=T-1}^0 p_\theta(\textbf{X}_t, \textbf{L}_t|\textbf{A}_0, \textbf{X}_{t+1}, \textbf{L}_{t+1})\label{eq:CrysVCD_diff}\\
    \text{where}\quad p(\textbf{A}_0)&= p(a_1)\prod_{i=1}^{n-1} p_\phi(a_{i+1}|a_1,\cdots ,a_i) 
\end{align}

Here, $p_\phi(\cdot)$ is a parametrized transformer model that performs autoregression for each elemental token $a_{i+1}$, which first generates the target atomic composition $\textbf{A}_0$ ($\textbf{A}_0$ is represented by the elemental tokens $(a_1,a_2,\cdots, a_n)$). Then, the second step of CrysVCD is equivalent to sampling $\textbf{X}$ and $\textbf{L}$ conditioned on $\textbf{A}_0$, thus sampling the joint distribution $p(\textbf{A}_0, \textbf{X}_0, \textbf{L}_0)=p(\textbf{A}_0)p(\textbf{X}_0,\textbf{L}_0|\textbf{A}_0)$.
Thus, $p_\theta(\cdot)$ is a parametrized E(3)-equivariant GNN learning the score function with one fewer degree of freedom, effectively performing crystal structure prediction (CSP) during the denoising process.
This CSP task is trained and implemented using the architecture of DiffCSP as a demonstration, but it can be achieved with other generative models as well. More details of the elemental language model in Stage I of CrysVCD are discussed in Methods.

Building on the two-step architecture, CrysVCD brings both efficiency and flexibility to crystal generation. The autoregressive elemental language model in Stage I requires a few inference steps (typically 4 or 5), orders of magnitude faster than the diffusion process (typically 1,000 denoising steps). This enables rapid pre-screening of chemically valid compositions before the computationally intensive CSP step (more details in Supplementary Information 3). Another core feature is the property guidance, which plays a key role in both steps. Target properties, such as $E_{\text{hull}}$, phonon stability, or functional attributes like thermal conductivity and dielectric constant, are embedded and used to condition both the composition generator and the denoising process. This allows CrysVCD to align material generation with desired physical or functional criteria, while remaining modular and extensible for future design goals.

\subsection*{Physics-inspired elemental language model embedding in CrysVCD}
As discussed above, a key innovation in CrysVCD is the two-step design that decouples composition generation from structural generation (Fig.\,\ref{fig1}). In this section, we detail Stage I for CrysVCD on the elemental language model. Mapping discrete vocabulary to continuous vectors, namely word2vec, is the foundation of the connection between languages and high-dimensional representation spaces in deep learning. 
In natural language processing, each word is typically assigned a randomly initialized embedding vector, which is learned by optimization through backpropagation across large corpora. However, for our elemental language model, where data scarcity is a significant challenge, we seek to incorporate prior chemical knowledge into the embedding design. 
Instead of random initialization, we engineer embeddings that reflect the periodic trends observed in materials science. For example, oxygen and sulfur exhibit similar valence behavior as chalcogens, yet differ in their metal affinities. 
The embedding also captures relationships and distinctions between multiple oxidation states of the same element, such as the valence diversity of manganese (Mn) from +2 to +7. Inspired by the SpookyNet machine learning potential~\cite{unke2021spookynet}, we employ the Aufbau principle to encode the electronic configurations of atoms and ions as the initial embeddings. 

\begin{figure}[!htbp]
    \centering\includegraphics[width=1.0\linewidth]{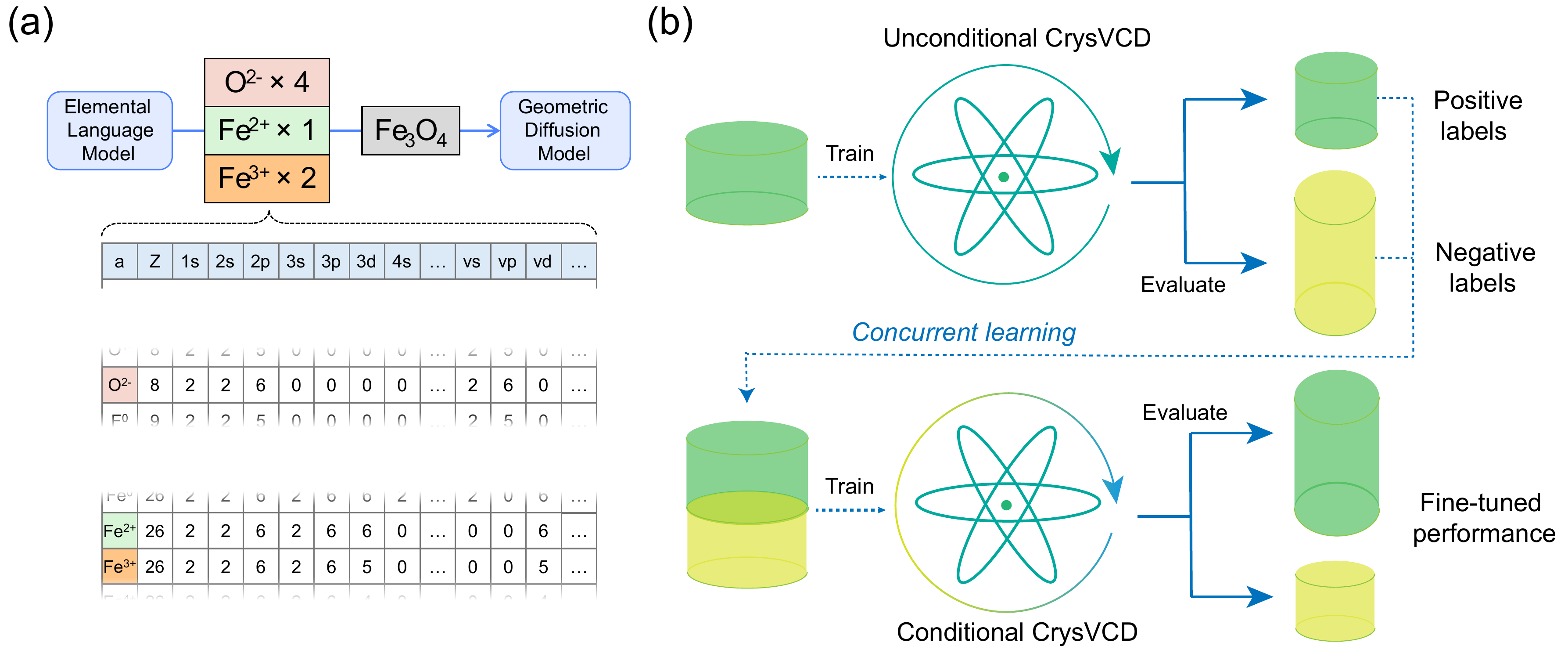}
    \caption{\textbf{Strategies of CrysVCD architecture and training}. \textbf{(a)} Tokenizer for the elemental language model as Stage I of CrysVCD. Chemical formulas are first parsed into valence-specific elemental tokens with electronic configurations of oxidation states. 
    These tokenized vectors are then passed into the geometric diffusion model (Stage II) for the structure generation of CrysVCD.
    \textbf{(b)} Concurrent learning scheme of CrysVCD for optimizing the generative quality. An unconditional CrysVCD model is initially trained on materials with positive labels, and generated structures are evaluated for target properties (e.g., stability metrics). 
    Based on these evaluations, structures are labeled as positive or negative and used to fine-tune a conditional CrysVCD model. This concurrent loop enables targeted generation with improved quality in terms of stability or other desired properties.}
    \label{fig2}
\end{figure}

Specifically, we construct a vocabulary containing common oxidation states of elements and decompose each material’s composition into element-valence pairs and their corresponding counts. Our decomposition algorithm successfully handles mixed-valence cases, such as the coexistence of \ce{Fe^{2+}} and \ce{Fe^{3+}} in magnetite (\ce{Fe3O4}), and explains the majority of compositions in the dataset. In this way, each word corresponds to a token that combines an element’s oxidation state with its atom count. As illustrated in Fig.\,\ref{fig2}(a), the model queries an electronic configuration table for element–valence pairs. This table encodes the atomic number, the occupancy of each electronic energy level, and the occupancy of the valence shell for each atom or ion. A learnable linear transformation then maps the electronic configurations to a continuous embedding vector, which are passed into a transformer model. During autoregressive generation, the model produces sequences of tokens specifying elements and their counts, which are combined into a chemically valid formula. This formula serves as input of the geometric diffusion model in Stage II. More details on the implementation of our elemental language model can be found in Methods and Supplementary Information 2.

\subsection*{Stability guided generation and fine-tuning in CrysVCD}
\begin{figure}[!htbp]
    \centering
    \includegraphics[width=0.9\linewidth]{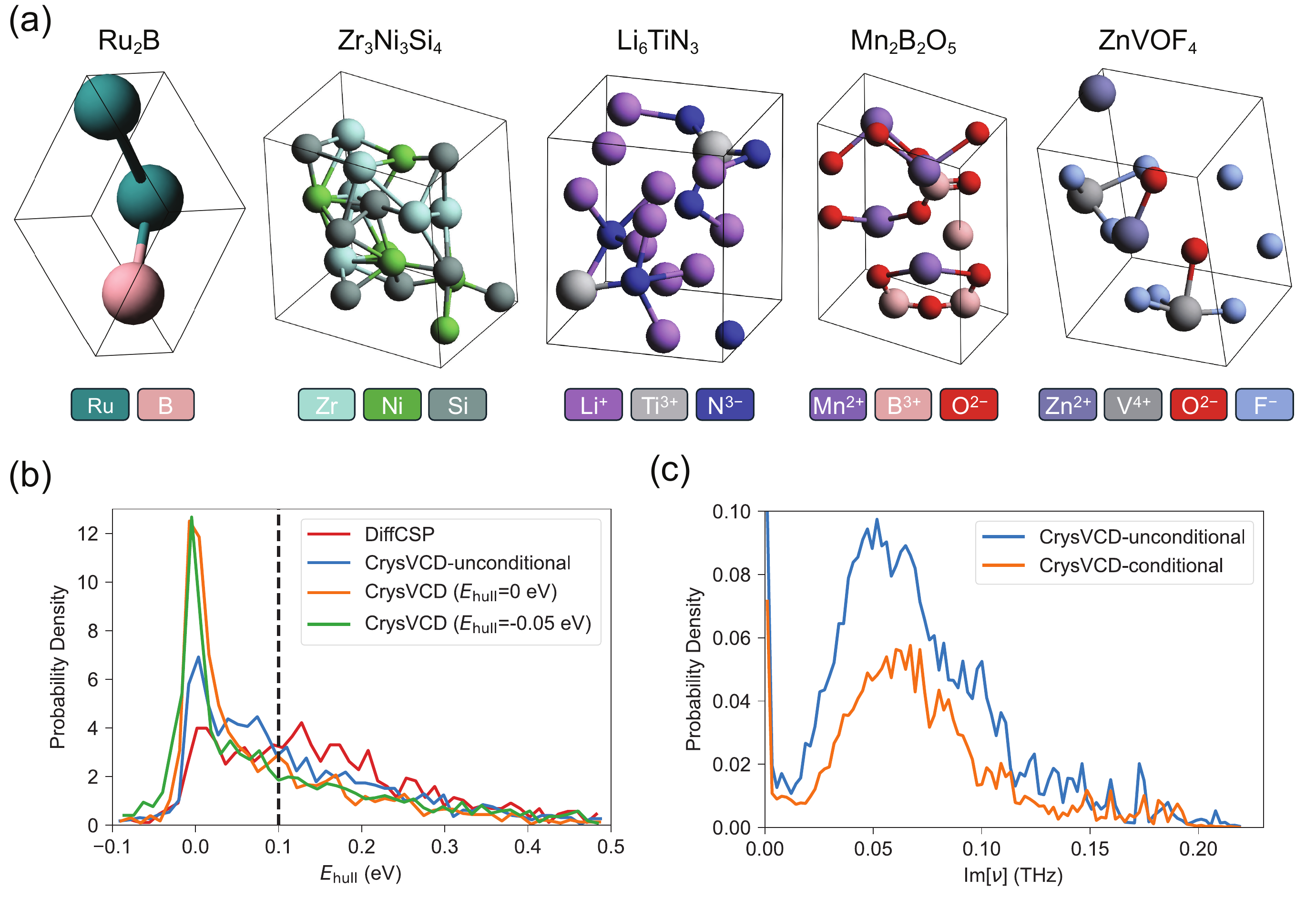}
    \caption{\textbf{Thermodynamic and phonon stability of generated crystals by CrysVCD}. \textbf{(a)} Examples of generated crystals from CrysVCD, including alloys (binary and ternary) and ionic compounds with oxidation states (ternary and quaternary) from left to right. 
    \textbf{(b)} Normalized probability distribution of the energy above hull ($E_{\rm hull}$) of crystal structures generated by different versions of CrysVCD and the previous generative model DiffCSP~\cite{jiao2023crystal}. CrysVCD fine-tuned with a $E_{\rm hull}$ threshold of 0 and -0.05 eV, and CrysVCD without conditional fine-tuning are included in comparison. \textbf{(c)} Normalized probability distribution of phonon imaginary frequencies of generated crystal structures by CrysVCD before and after conditional fine-tuning with labels of phonon stability evaluated by the MatterSim-v1.0.0-5M model~\cite{zeni2025generative}. }
    \label{fig3}
\end{figure}

Building on the CrysVCD architecture, we present examples of generated crystals in Fig.\,\ref{fig3}(a), including alloys and ionic compounds with varying chemical complexity.
The oxidation states for all elements are all labeled in Fig.\,\ref{fig3}(a), and all generated crystals are guaranteed to be valence-balanced by construction.
To evaluate the structural quality of the generated crystals, we assess two stability metrics, (1) the energy above hull ($E_{\rm hull}$) and (2) the phonon imaginary frequencies of the generated structures.
Structures with low $E_{\rm hull}$ and absence of imaginary phonon modes are generally regarded as indicators of thermodynamic and dynamic stability, respectively. 
A key feature of CrysVCD compared to other generative approaches lies in its explicit incorporation of stability-guided learning. 
Other materials generative models often operate in a purely data-driven manner, learning directly from known structure databases composed of stable compounds. This introduces an intrinsic bias into the generative process that limits their ability to generalize beyond the training distribution. 

To overcome this limitation, we adopt a concurrent learning framework that integrates unconditional generation with stability-conditioned fine-tuning, as shown in Fig.\,\ref{fig2}(b). 
The process begins with training an unconditional CrysVCD model on the MP-20 dataset to learn general crystal representations. This model is then used to generate a diverse set of hypothetical compounds, which are explicitly evaluated using high-throughput stability screening.
These results are labeled as either stable positive (green) and unstable negative (yellow) and used to construct a new dataset.
These newly labeled samples are then aggregated into a training dataset (dotted blue line) for a second-step conditional CrysVCD model, which is trained to generate structures conditioned on stability labels. 
This closed-loop feedback pipeline enables active learning in CrysVCD,  where the model not only generates but also learns from its own outputs. By incorporating both stable and unstable generations, CrysVCD gains insight into structural features that lead to instability and can adapt accordingly.

We benchmark the stability of materials generated by CrysVCD in Fig.\,\ref{fig3} in terms of stability metrics. Even without conditional fine-tuning, CrysCVD exhibits lower $E_{\rm hull}$ distribution compared to the existing DiffCSP model, as shown in Fig.\,\ref{fig3}(b), despite the fact that the diffusion model of CrysVCD is DiffCSP. This is attributed to Stage I of CrysVCD, which ensures balanced chemical valence for generated crystals, and demonstrates the generalizability of CrysVCD to improve other generative models. 
To further evaluate the stability of the generated crystals, we use MatterSim to evaluate the $E_{\rm hull}$ and phonon frequencies, and use these labels to fine-tune the CrysVCD model. The stability fine-tuned CrysVCD model generates substantially lower $E_{\rm hull}$ distribution than the unconditional CrysVCD model. Moreover, the fine-tuned CrysVCD generates less phonon imaginary frequencies than the unconditional CrysVCD, as shown in Fig.\,\ref{fig3}(c). The portion of phonon-stable configurations (without any imaginary frequency) in the generated crystals increases from 49\% (unconditional) to 68\% (fine-tuned). These results demonstrate that CrysVCD can effectively generate phonon stable structures.

\subsection*{Generating high-performance functional materials with CrysVCD}

Having established the architecture and stability-guided learning framework of CrysVCD, we now demonstrate its capability to generate functional materials tailored for real-world applications. Specifically, we focus on two technologically important categories: materials with intrinsically high thermal conductivity and those exhibiting high dielectric constants (high-$\kappa$), both of which are crucial for next-generation microelectronic systems.
Materials with high phonon thermal conductivity ($k_{\text{ph}}$) but electrically insulating are vital for thermal management in microelectronics and energy systems \cite{zheng2021advances}. 
While high-throughput screening have identified promising candidates \cite{plata2017efficient,pota2024thermal,li2025probing}, finding stable, intrinsically high $k_{\text{ph}}$ remains challenging. Diamond leads in $k_{\text{ph}}$ but has high cost, and other recent generative efforts are limited to carbon allotropes\cite{guo2025generative}. 

\begin{figure}[!htbp]
    \centering
    \includegraphics[width=\linewidth]{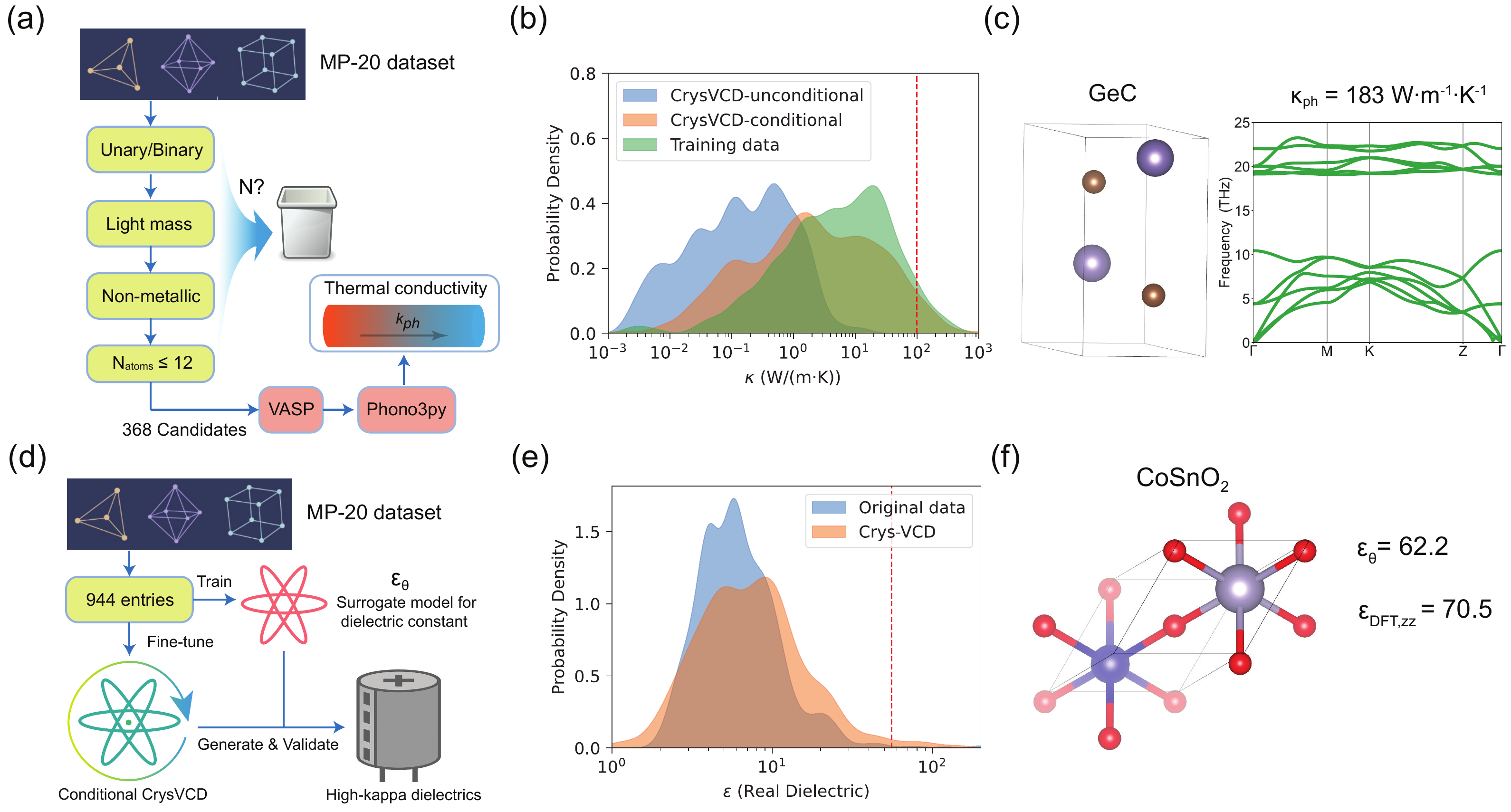}
    \caption{\textbf{Generation of high-performance functional materials with CrysVCD}. \textbf{(a)} Workflow of first-principles calculations to generate a thermal conductivity dataset considering phonon anharmonicity. \textbf{(b)} Probability distribution of the thermal conductivity of materials generated by the unconditional CrysVCD model and CrysVCD fine-tuned to maximize thermal conductivity. The distribution of high thermal conductivity materials from the training dataset selected by human intuition is shown as a reference. The red dashed line denotes $k_{\text{ph}} = 100 ~\SI{}{W\cdot m^{-1}\cdot K^{-1}}$ as the generative condition. \textbf{(c)} First-principles calculations of the phonon band structure of a high thermal conductivity material GeC, generated by the CrysVCD model fine-tuned to maximize thermal conductivity. 
    \textbf{(d)} Workflow for conditional CrysVCD generation on dielectric constant. A surrogate model for static dielectric constant ($\varepsilon$) is trained on 944 MP-20 entries, then used to guide CrysVCD toward generating structures with high $\varepsilon$. \textbf{(e)} Probability distribution of $\varepsilon$ for generated materials (orange) versus original MP-20 dataset (blue), showing a rightward shift indicating successful targeting of high-$\varepsilon$ candidates. The red dashed line denotes $\varepsilon$ = 50 as the generative condition. \textbf{(f)} A new high-$\kappa$ candidate generated by CrysVCD CoSnO$_2$. The surrogate model predicts a dielectric constant $\varepsilon_\theta = 62.2$, with DFT verification yielding $\varepsilon_{\mathrm{DFT}, zz} = 70.5$.}
    \label{fig4}
\end{figure}

To discover more high-$k_{\text{ph}}$ materials, we fine-tune CrysVCD using a screened subset of MP-20 dataset guided by phonon transport knowledge. As shown in Fig.\,\ref{fig4}(a), we prioritize materials that are likely to exhibit high phonon group velocities and long phonon lifetimes \cite{togo2015distributions}, by focusing on materials with lighter elements, strong covalent or mixed bonding, simple unit cells (< 12 atoms), which can increase group velocities and reduce phonon scattering. Moreover, we exclude metallic compounds and gas-phase species to focus on phonon-dominated heat transport. 

The resulting screened dataset serves both as the training set for conditional CrysVCD, as well as a benchmark for assessing its ability to emulate expert-driven design.
After fine tuning the conditional generative model for thermal conductivity targets, CrysVCD shifts its generative distribution toward high-$k_{\text{ph}}$ candidates, as shown in Fig.\,\ref{fig4}(b). Compared to the base model, the conditioned version generates structures with $k_{\text{ph}}$ values aligned with expert-curated materials.

Among the generated compounds, we highlight hexagonal GeC with high thermal conductivity $k_{\text{ph}} = 183~\SI{}{W\cdot m^{-1}\cdot K^{-1}}$. Notably, this material does not appear in the MP-20 dataset. 
Its crystal structure and phonon band structure are shown in Fig.\,\ref{fig4}(c). Combined with a with a wide bandgap of 2.34 eV, GeC is a promising candidate for thermal management applications. The large acoustic-optical phonon gap shown suppresses phonon scattering, contributing to its high thermal conductivity.
Interestingly, given the simplicity of GeC, it is present in the full Materials Project database, yet has not been systematically studied for thermal transport. CrysVCD's ability to generate GeC without prior exposure highlights its potential for uncovering promising functional materials.

To further validate the efficacy of CrysVCD, we conduct another study on high-$\kappa$ dielectrics, i.e., materials with high dielectric constant $\varepsilon$. 
These materials are of significant technological interest for applications in capacitors, transistors, and other microelectronic device. Due to the high computational cost of DFT-based optical properties, it is impractical to evaluate each generated structure using first-principle methods. 
To address this bottleneck, we leverage an existing database and train another machine learning-based surrogate model that can provide rapid and reasonably accurate predictions of $\varepsilon$.
As shown in Fig.\,\ref{fig4}(d), we begin by collecting 944 entries from the Materials Project within the MP-20 dataset, which contains DFT-computed dielectric data. 
This dataset is used for both fine-tuning the CrysVCD model for conditional generation and training the surrogate model for predicting the dielectric constant.
The surrogate model is implemented as an $E(3)$-equivariant graph neural network \cite{reiser2022graph,geiger2022e3nn}, adapted from a GNNOPT architecture for optical property prediction \cite{hung2024universal}. Once trained, the GNNOPT model is integrated into the conditional generation loop of fine-tuned CrysVCD to rapidly evaluate and guide the selection of candidate materials with high $\varepsilon$. The predictive performance of the $E(3)$-equivariant surrogate model is shown in Supplementary Information 3.

The resulting distribution of $\varepsilon$ for generated materials, shown in Fig.\,\ref{fig4}(e), exhibits a clear shift toward higher $\varepsilon$ values compared to the original MP-20 dataset. During conditional generation, we explicitly set the target property to $\varepsilon = 50$, guiding CrysVCD to prioritize candidates above this threshold. 
As a result, a substantial number of generated materials achieve $\varepsilon > 50$, in contrast to the original dataset, where such high-dielectric entries are rare. This demonstrates the effectiveness of CrysVCD to steer the generative process toward chemically valid, high-$\varepsilon$ candidates. 
Among the top-performing candidates, we identify CoSnO$_2$, which is not present in the Materials Project database. It exhibits a high dielectric constant with $\varepsilon_\theta = 62.2$ predicted by the surrogate model and a principal component value of $\varepsilon_{\mathrm{DFT}, zz} = 70.5$ verified by DFT, as shown in Fig.\,\ref{fig4}(f).
However, it should be noted that these results rely on a surrogate model trained on DFT-computed dielectric data, which may still deviate from experimental values. Improving surrogate accuracy and bridging the gap between DFT and real-world behavior remain important directions for future work.

\section*{Discussion}
Our work highlights the potential of chemical composition as a natural modality for generative modeling in materials science, which is widely supported in predictive models~\cite{ma2023topogivity,ma2025learning}. By first generating valid chemical compositions and then producing their crystal structures, our work helps yield stable candidates that are consistent with valence balance and bonding principles.
Unlike purely data-driven approaches, our framework is explicitly guided by physical and chemical constraints. The chemical composition generation is explicitly guided by charge neutrality constraints, and element embeddings are initialized using electronic configurations. This design enables the model to capture periodic trends, accommodate multiple oxidation states, and handle complex chemistry such as mixed-valence compounds.

To address the scarcity and uneven distribution of experimental data, such as thermal conductivity or dielectric constant, we integrate the generation pipeline with machine-learning-accelerated predictive models. These models are not only applied for efficient \textit{post hoc} evaluation of generated candidates, but also provide synthetic property labels that can be used to steer the generative model in subsequent rounds, or accommodate data augmentation from human expertise. This generation-prediction concurrence transforms the model into an active explorer of the chemical space, extrapolating beyond the support of the original dataset and proposing materials with novel compositions and desirable properties.

Nevertheless, the performance of CrysVCD, is still constrained by the MP-20 training data, which excludes materials containing lanthanides, actinides, molecular crystals, and other complex motifs such as zeolites and metal-organic frameworks (MOFs). Within the remaining chemical space in MP-20 dataset, although our valence decomposition covers 97.6\% of the materials, there are potentially high-value materials with exotic oxidation states being overlooked by our model.
Future work may extend our model to a broader chemical and structural space, with larger and more diverse training data. Ultimately, CrysVCD is designed as a modular and extendable plugin architecture that can be readily integrated into existing generative frameworks to enforce chemically valid, charge-balanced generation.

\section*{Methods}
\subsection*{Chemical formula generation}
We define a set of \textit{valid atom types} $A$ by excluding noble gases, lanthanides, actinides, and other radioactive species from the periodic table. We manually compiled a list of \textit{commonly observed non-zero oxidation states} in ionic compounds and assigned an additional zero valence to those atom types that typically appear in alloys, forming a set of \textit{valid oxidation states} $\bm V=\{V(a):a\in A\}$.
Using this oxidation state list, we developed a valence labeling algorithm for the compounds containing only valid atom types in the MP-20 subset of the Materials Project~\cite{jain2013commentary}. Each material's \textit{valence assignment} is represented as a set of tuples $\{(a_i^{v_i}, c_i)\}_i$, where $a^v$ denotes the atom type $a$ in valence state $v$, and $c$ denotes the number of atoms of type $a$ and valence $v$ in the unit cell. A valid assignment must satisfy the charge neutrality condition $\sum c_iv_i=0$.

For a material with a unit-cell stoichiometry \ce{A_xB_y\cdots C_z}, we classify it as an alloy if $0\in V(a)$ for all $a\in\{\ce{A},\ce{B},\cdots,\ce{C}$\}, and label it as $\{(\ce{A^0},x),(\ce{B^0},y),\cdots,(\ce{C^0},z)\}$. Otherwise, it is treated as an ionic compound. For ionic materials, we perform a breadth-first search over all non-zero valence combinations from $\bm V$ to identify a valid assignment. Elements may be assigned multiple valence states within the same structure; for example, \ce{Fe3O4} is labeled as $\{(\ce{Fe^{2+}}, 1), (\ce{Fe^{3+}}, 2), (\ce{O^{2-}}, 4)\}$. To improve efficiency, we apply pruning based on plausible valence ranges in $\bm V$. Additionally, we impose \textit{comproportionation constraints} to avoid infeasible configurations: no element may simultaneously exhibit both positive and negative oxidation states, and if three oxidation states $u<v<w$ appear in $V(a)$, then coexistence of $a^{u},a^{w}$ is prohibited. Materials that fail to obtain a valid valence assignment or contain invalid elements are excluded. The remaining successfully labeled materials form the dataset \textbf{MP-20-valence}, which has 8,299 alloys and 7,491 ionic compounds in its training partition.

We adopt the GPT-2~\cite{radford2019language} implementation on the Huggingface platform \cite{wolf2019huggingface} for our \textit{chemical formula transformer}, and design a domain-specific embedding scheme tailored to the valence assignment representation in MP-20-valence. For a count $c$, ranging from 0 to 20 in all MP-20-valence entries, we embed it into 21 learnable word embeddings $\bm h_c\in\R^{d}$. For a valence-labeled atom type $a^v$ we define a vocabulary of 217 unique tokens in $\{a^v:a\in A, v\in V(a)\}$, augmented with two special tokens: \verb|<START>| and \verb|<END>|. They are first embedded into 219 learnable word embeddings $h_{a^v}^{(1)}\in\R^{d}$. Furthermore, inspired by the electronic embedding proposed in SpookyNet~\cite{unke2021spookynet}, we encode the electron counts of each orbitals in their full electronic configuration and the valence shell configuration for each $a^v$ token by a linear transformation into a vector $\bm h_{a^v}^{(2)}\in\R^{d}$, while special tokens are mapped to zero vectors of the same dimensionality. As a result, each $(a^v, c)$ pair is embedded as the combination $\bm h_{a^v}^{(1)}+\bm h_{a^v}^{(2)}+\bm h_c\in\R^d$. A valence assignment of $l$ pairs will be converted to a sequence, padded with $(\verb|<START>|,0)$ and $(\verb|<END>|,0)$ at its head and tail, and transformed into an embedding sequence of $\R^{(L+2)\times 41}$, which serves as the word embedding of the transformer.

The transformer performs next-token prediction in an autoregressive way. The next-token probability is predicted as a concatenated logit $[\bm y_{a_v}||\bm y_c]\in\R^{240}$, where $\bm y_{a_v}\in\R^{219}$ corresponds to the logit of valence-labeled element tokens, and $\bm y_c\in\R^{21}$ corresponds to the logit of count tokens. In the training stage, we augment each training data entry by at most 3 times with random permutations of the valence assignment tuples. Two transformers for alloys and ionic compounds are separately trained with the total cross-entropy loss of the two logits in a teacher-forcing scheme~\cite{vaswani2017attention}. In the inference stage, the sequence generation starts with $(\verb|<START>|,0)$, extends by sampling the next token $(a^v,c)$ with the highest probability, and ends when it reaches the maximum length of 10 or generates \verb|<END>| for $a^v$. We append a filter to the transformer for ionic compounds to ensure that only charge-balanced generation results are passed to the diffusion model. The initial atom-type condition $\bm A_0$ of the diffusion model in \autoref{eq:CrysVCD_diff} is compiled from $a_i$ and $c_i$ of a valid generated sequence.

\subsection*{Property-guided conditional crystal generation}
Our model supports two types of properties as the generation condition. The first type is \textit{binary properties}, taking values in $\{0,1\}$, such as phonon stability. The second type is continuous properties, represented as real-valued scalars, including properties like $E_\text{hull}$ and thermal conductivity. Each property has an \textit{adapter channel} that transforms its value into an embedding vector $\bm h_p$ for the chemical formula transformer and the crystal diffusion model, respectively. A binary property is embedded by a word-embedding layer of size 2, while a continuous property is transformed by a linear mapping. When conditioned on multiple properties, all property embeddings are added together.

The chemical formula transformer adds its own $\bm h_p$ to the atom-valence and count embedding of the \verb|<START>| during the training on property-labeled datasets and conditioned generation. The crystal diffusion model adds its own $\bm h_p$ to each initial atomic feature in its score estimator. One unified score estimator for both alloys and ionic compounds is trained with classifier-free guidance~\cite{ho2022classifier} throughout this work. We adopted a guidance weight $w=2$ to scale the distribution conditioned on $\bm h_p$. It replaces $p(\bm X_t,\bm L_t)$ in \autoref{eq:CrysVCD_diff} during the conditional denoising process. 

\begin{equation}\label{eq:cfg}
\begin{aligned}
p_w(\bm X_t,\bm L_t\mid \bm h_p) & \propto p(\bm h_p \mid \bm X_t,\bm L_t)^w p (\bm X_t,\bm L_t) \\
& \propto\left(\frac{p(\bm X_t,\bm L_t \mid \bm h_p)}{p(\bm X_t,\bm L_t)}\right)^w p(\bm X_t,\bm L_t) \\
& \propto p(\bm X_t,\bm L_t \mid \bm h_p)^w p(\bm X_t,\bm L_t)^{1-w}
\end{aligned}
\end{equation}

The conditional score corresponding to \autoref{eq:CrysVCD_diff} is obtained by taking gradients of the logarithm with respect to $\bm X_t$ and $\bm L_t$. 
$$
\nabla \ln p_w\left(\bm X_t,\bm L_t \mid \bm h_p\right)=w \nabla \ln q_\theta\left(\bm X_t,\bm L_t \mid \bm h_p\right)+(1-w) \nabla \ln q_\theta\left(\bm X_t,\bm L_t\right) .
$$
where the conditional score is estimated by the score model with property embeddings $\bm h_p$, and the unconditional score is obtained by replacing $\bm h_p$ with a zero vector.

\subsection*{MLIP and DFT calculations}
The $E_{\rm hull}$ is evaluated using the MatterSim-v1.0.0-1M model with the convex hull information from the Materials Project database~\cite{jain2013commentary} using the pymatgen interface~\cite{ong2013python}. The atomic configuration is relaxed until the max atomic force is less than 0.01 eV/\AA \,before the $E_{\rm hull}$ evaluation. For phonon stability, we use the MatterSim-v1.0.0-5M model to evaluate the phonon band structure. The atomic configuration is first relaxed until the maximum atomic force is less than 0.003 eV/\AA \,using the BFGS algorithm~\cite{dai2002convergence}. 
Note that higher accuracy settings are adopted due to the high requirements for phonon calculations. If the relaxation does not converge within 200 steps, the structure is considered unstable. The force constant matrix of the system is then calculated by the finite difference method with a differential step size of 0.1 \AA.  Phonon spectrum is evaluated using the phonon module of Atomic Simulation Environment~\cite{larsen2017atomic} with a supercell size of at least 20 \AA \,in each direction and a k-point mesh of $8\times 8\times 8$. Phonon frequencies at all k points in each configuration are included in the statistics in Fig.~\ref{fig3}b. If an imaginary frequency larger than $0.01$ THz appears in the computed phonon band structure, the structure is labeled unstable.

For the calculation of thermal conductivity, we use the phono3py package \cite{phonopy-phono3py-JPCM,phonopy-phono3py-JPSJ}, which requires second-order and third-order force constants to perform anharmonic phonon scattering calculations.
We supply these force constants using MatterSim-v1.0.0-5M for screening, and DFT results for benchmarking and verification of generated high-thermal conductivity materials.
DFT calculations are conducted using the Vienna Ab initio Simulation Package (VASP)\cite{kresse1996efficient}. The projector-augmented wave (PAW) method\cite{blochl1994projector,kresse1999ultrasoft} is employed, with exchange-correlation effects described by the Perdew-Burke-Ernzerhof (PBE) formulation of the generalized gradient approximation (GGA)\cite{perdew1996generalized}. The plane-wave cutoff energy is set as $1.2\times\text{max(ENMAX)}$ for sufficient convergence. 
Calculations are performed on a K-point mesh centered at the $\Gamma$ point with resolved value $k_{\text{mesh}}=0.03\cdot 2\pi$/$\mathring{\text{A}}$ for each structure.
Additionally, our data analysis on DFT results utilizes the VASPKIT package\cite{wang2021vaspkit}.
For both MLIP and DFT calculations of force constants, we construct supercells using the finite-displacement approach under the frozen phonon method.

\section*{Competing interests}
The authors declare that a patent application has been filed relating to the material described in this manuscript.

\section*{Data and code availability}
The data and code used in this study are available at \url{https://github.com/vipandyc/CrysVCD}.

\section*{Acknowledgements}
The authors thank G Daras, BR Ortiz, Q Yan and RJ Cava for helpful discussions.
M.C. acknowledges the support of U.S. Department of Energy, Office of Science, Basic Energy Sciences, award No. DE-SC0021940.
W.L. and H.J.K. acknowledge the support of the U.S. Department of Energy, Office of Science, Office of Advanced Scientific Computing, Office of Basic Energy Sciences, via the Scientific Discovery through Advanced Computing (SciDAC) program. 
H.T. acknowledges the support of the Mathworks Engineering Fellowship.
Y.C. is partly supported by the Scientific User Facilities Division, Office of Basic Energy Sciences, U.S. Department of Energy, under Contract No. DE-AC0500OR22725 with UT Battelle, LLC. A portion of the DFT calculations used computing resources made available through the VirtuES project, funded by the Laboratory Directed Research and Development program and Compute and Data Environment for Science (CADES) at Oak Ridge National Laboratory.
M.L. acknowledges the support of National Science Foundation (NSF) ITE-2345084, and the support from R. Wachnik.

\bibliography{refs_intro.bib} 

\end{document}


\maketitle
\flushbottom
\tableofcontents
\thispagestyle{empty}

\section{Preparation of dataset}
\subsection{Statistics of crystal structure and composition}

In this section, we present key statistics regarding the structural prototypes and chemical compositions in the dataset to provide a comprehensive understanding of the data distribution used for CrysVCD.

\begin{figure}[h]
    \centering
    \includegraphics[width=\textwidth]{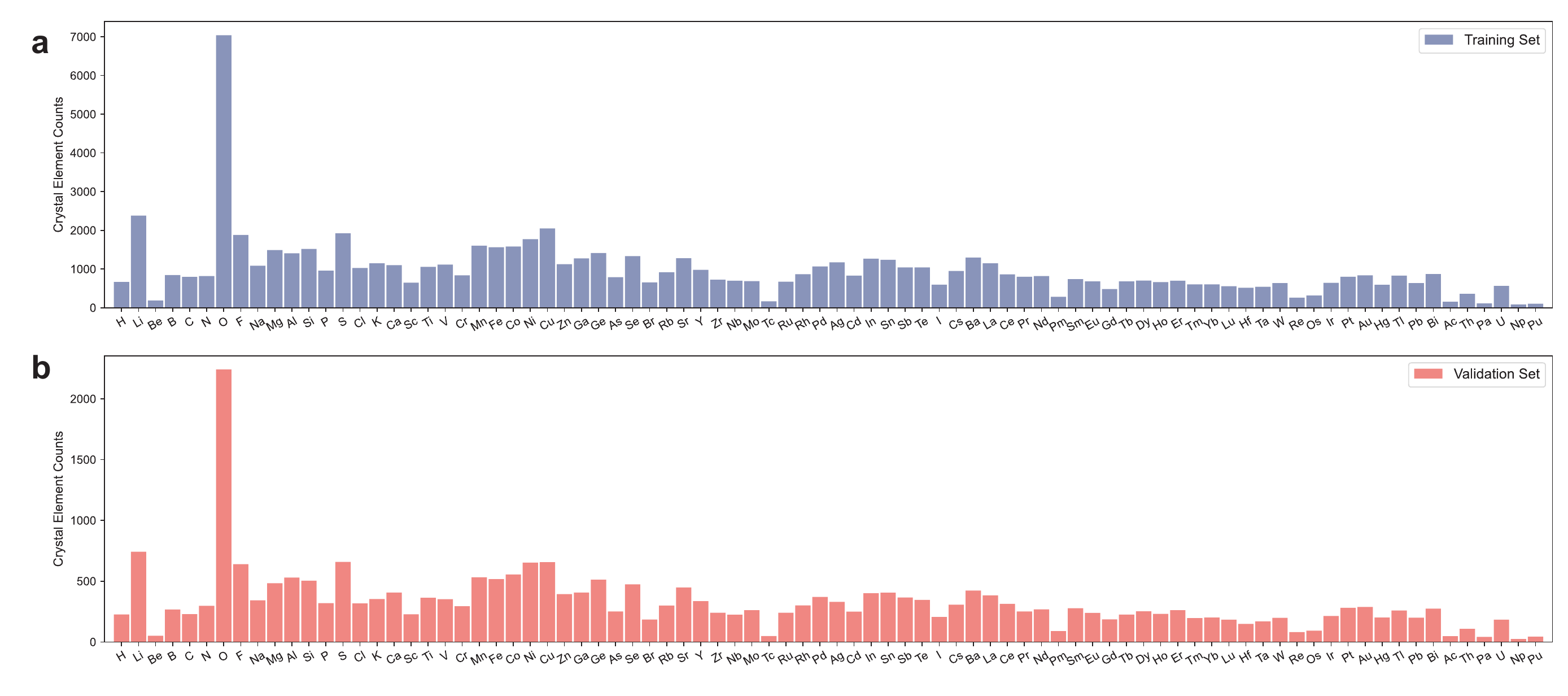}
    \caption{Elemental distribution statistics among the training and validation datasets. \textbf{a.} Elemental composition of crystals in the training dataset. \textbf{b.} Elemental composition of crystals in the validation dataset.}
    \label{fig:elements}
\end{figure}

Regarding chemical diversity, Fig.\,\ref{fig:elements} illustrates the dataset's elemental distribution, covering a total of 84 distinct elements. Fig.\,\ref{fig:elements}(a) and (b) highlight the frequency of each element within the training and validation datasets, respectively. Common elements, including O, Cu, Li, F, and S, exhibit substantial representation, while rare-earth and heavy elements appear comparatively infrequently. Notably, noble gases are completely absent from the dataset due to their limited propensity for crystal formation.

\begin{figure}[h]
    \centering
    \includegraphics[width=\textwidth]{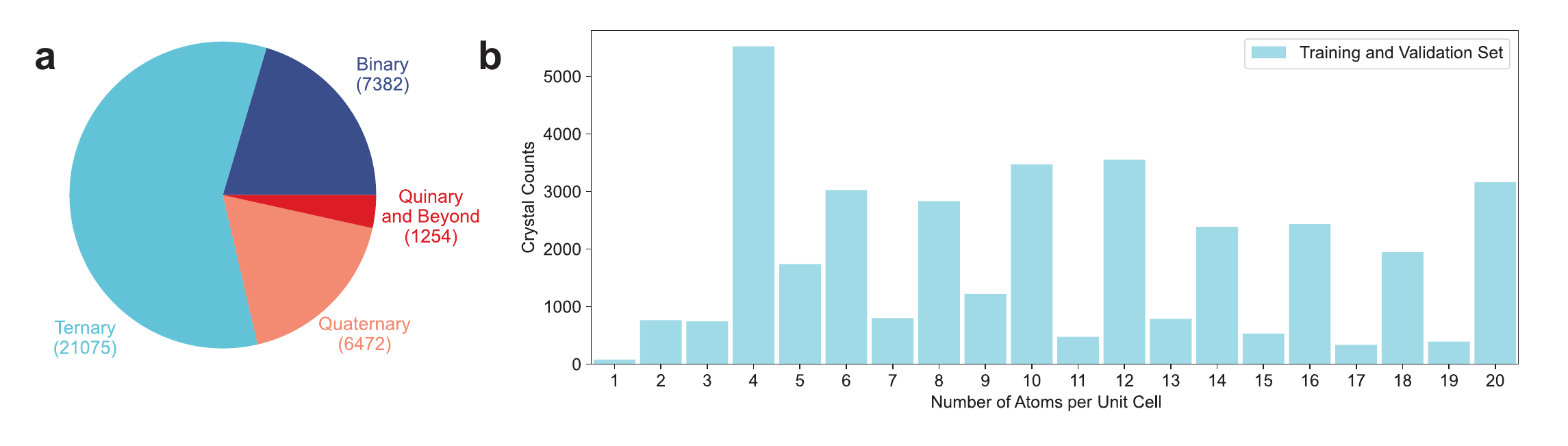}
    \caption{Structural statistics of the entire dataset, including both training and validation ones. \textbf{a.} Distribution of crystal compositions, most of which are among binary to quinary systems. \textbf{b.} Distribution of the number of atoms per unit cell.}
    \label{fig:crystals}
\end{figure}

Fig.\,\ref{fig:crystals} provides an overview of crystal structure diversity. Panel (a) depicts the distribution of crystal compositions across the entire dataset--including both training and validation sets. The majority of crystals fall within the binary to quinary range, with only a minimal number of crystals consisting of more than five elements. Among all crystals in the dataset, ternary crystals are notably predominant, with more than 20,000 occurrences. Panel (b) details the distribution of atoms per unit cell, ranging from 1 to 20 atoms.

\subsection{List of chemical valence}
\begin{figure}[!htbp]
    \centering
    \includegraphics[width=\linewidth]{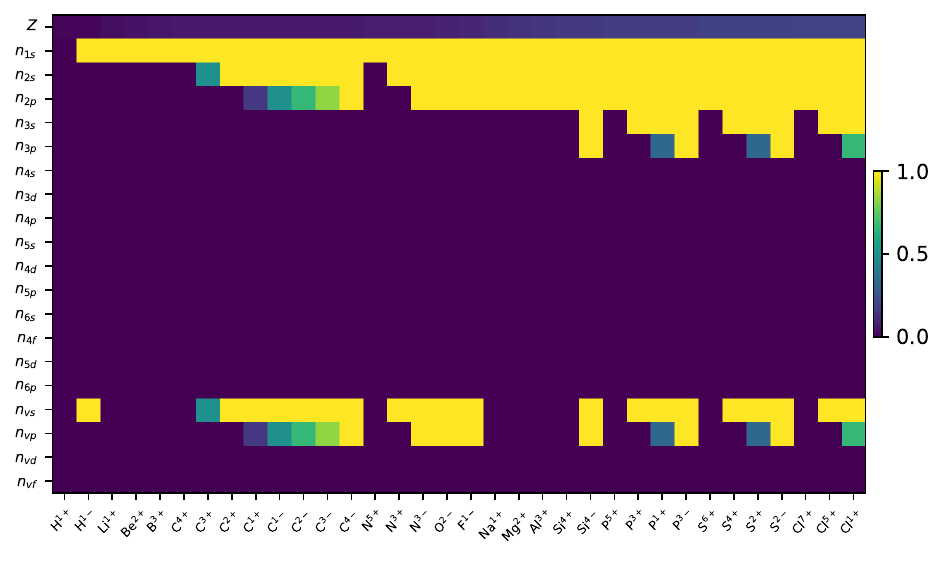}
    \caption{Electronic configuration embeddings of ions in the first–third rows of the periodic table (H-Cl). The $x$ axis lists the atom types and the $y$ axis stands for each entry in the embedding vector including the nuclear charge and the electronic shell occupancy under the aufbau principle.}
    \label{fig:iec_1}
\end{figure}

\begin{figure}[!htbp]
    \centering
    \includegraphics[width=\linewidth]{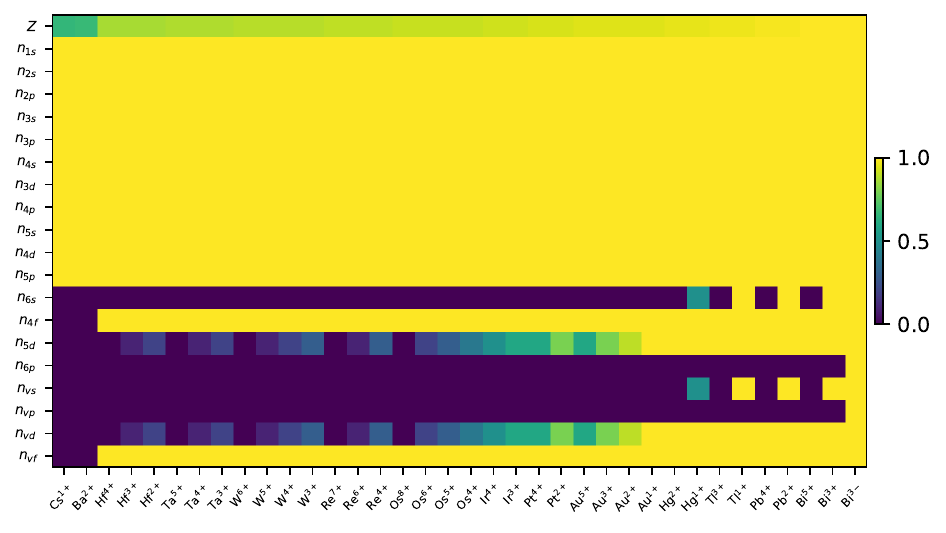}
    \caption{Electronic configuration embeddings of ions in the sixth row of the periodic table (Cs-Bi). The $x$ axis lists the atom types and the $y$ axis stands for each entry in the embedding vector including the nuclear charge and the electronic shell occupancy under the aufbau principle.}
    \label{fig:iec_4}
\end{figure}

\begin{figure}[!htbp]
    \centering
    \includegraphics[width=\linewidth]{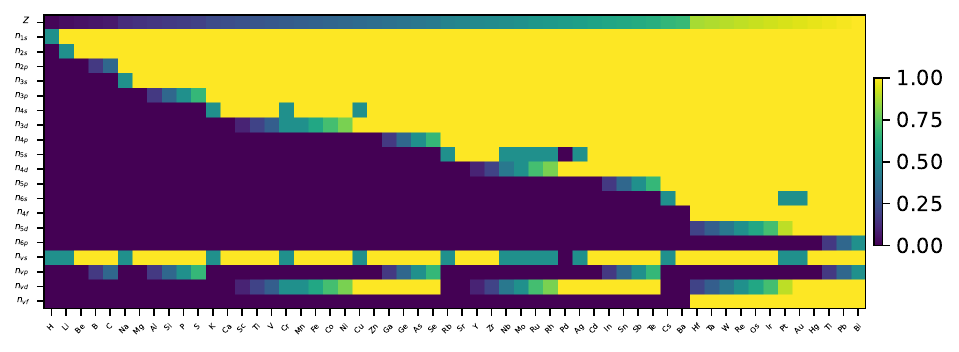}
    \caption{Electronic configuration embeddings of elements in alloys. Including nonmetallic elements like H, B, C and common metallic elements. The $x$ axis lists the atom types and the $y$ axis stands for each entry in the embedding vector including the nuclear charge and the electronic shell occupancy under the aufbau principle.}
    \label{fig:aec}
\end{figure}

Below, we list the electronic configuration embeddings for the chemical formula transformer in Fig.\,\ref{fig:iec_1}-Fig.\,\ref{fig:iec_4} for ions and Fig.\,\ref{fig:aec} for elements in alloys. The physical meaning of each entry is labeled at the $y$-axis, where $Z$ is the atomic number (nuclear charge), $n_\text{1s}$ is the number of 1s electrons in its ground-state electronic configuration, and $n_\text{vs}$ is the number of s electrons in valence shells. The numerical value of each entry is normalized by the maximum value in our scope of the periodic table.

\begin{figure}[!htbp]
    \centering
    \includegraphics[width=\linewidth]{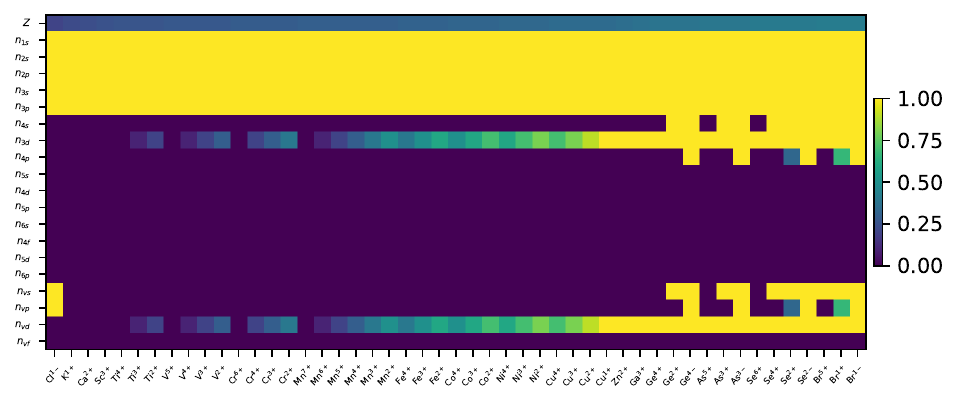}
    \caption{Electronic configuration embeddings of ions in the third-fourth rows of the periodic table (Cl-Br). The $x$ axis lists the atom types and the $y$ axis stands for each entry in the embedding vector including the nuclear charge and the electronic shell occupancy under the aufbau principle.}
    \label{fig:iec_2}
\end{figure}

\begin{figure}[!htbp]
    \centering
    \includegraphics[width=\linewidth]{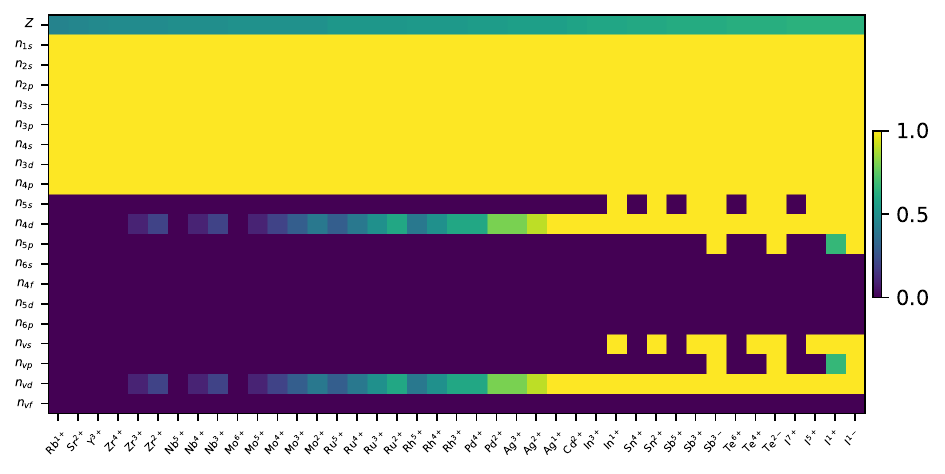}
    \caption{Electronic configuration embeddings of ions in the fifth row of the periodic table (Rb-I). The $x$ axis lists the atom types and the $y$ axis stands for each entry in the embedding vector including the nuclear charge and the electronic shell occupancy under the aufbau principle.}
    \label{fig:iec_3}
\end{figure}

\newpage
\section{Architecture for CrysVCD}
\subsection{Elemental language model}

\begin{figure}[!htbp]
    \centering
    \includegraphics[width=0.5\linewidth]{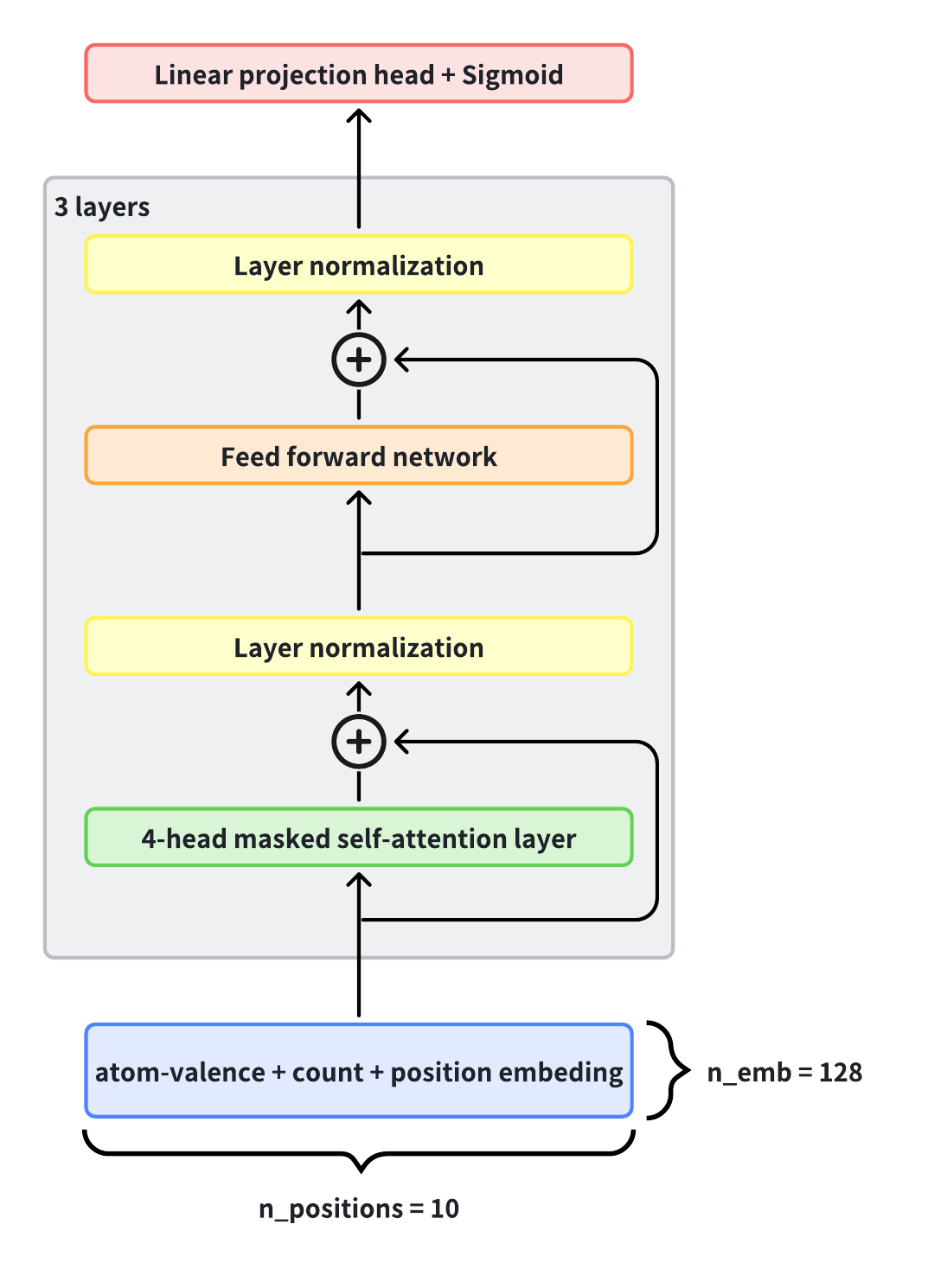}
    \caption{The model architecture of the chemical formula transformer used in CrysVCD.}
    \label{fig:element_GPT}
\end{figure}

The detailed architecture of our chemical formula transformer is shown in Fig.\,\ref{fig:element_GPT}, whose implementation is provided by the HuggingFace version of GPT-2. Due to the small scale of training data, we cut down the size of the default options to reduce the computational cost and avoid potential overfitting. The dimension of the token embedding \verb|n_emb| is set to 128. The \verb|n_position|, corresponding to the maximum sequence length, is set to 10 based on the statistics of our chemical formula dataset. The number of transformer layers \verb|n_layer| is set to 3, and the multi-head self-attention mechanism uses \verb|n_head|=4 heads. We adopt HuggingFace's default values for all other architectural hyperparameters. Both pretraining and fine-tuning use a learning rate of $10^{-3}$ in the Adam optimizer, with a fixed epoch number of 100.

\subsection{Diffusion module for crystal structure prediction}
Our workflow allows for any crystal structure prediction algorithm, including genetic algorithms, particle-swarm algorithms, deep generative models like variational auto-encoders, large language models, as well as the diffusion model, to build the structure from a chemically valid chemical formula generated by the chemical formula transformer. As a showcase, we utilize the code from the official implementation of the DiffCSP model on \href{(https://github.com/jiaor17/DiffCSP)}{Github} as the crystal structure prediction module in our material design workflow, and set the same architecture and training hyperparameters as the default of the MP-20 task in the DiffCSP codebase. We keep the pretrained model with the best loss on the validation set in 1000 epochs at a learning rate of $10^{-3}$ in the Adam optimizer. When fine-tuning on phonon stability or energy above the hull, the learning rate is reduced to $5\times 10^{-4}$, and further reduced to $10^{-4}$ during thermal conductivity or dielectric constant fine-tuning. Our evaluation is also based on the fine-tuned model with the best loss on the validation set in 1000 epochs.

\newpage
\section{Additional results for CrysVCD}
Below, we show additional results related to CrysVCD in the conditional generation tasks, including computational overhead, stability evaluation, and surrogate model prediction on the dielectric constant.

\subsection{Computational overhead of the chemical formula transformer on the material generation}
\begin{figure}[!htbp]
    \centering
    \includegraphics[width=0.85\linewidth]{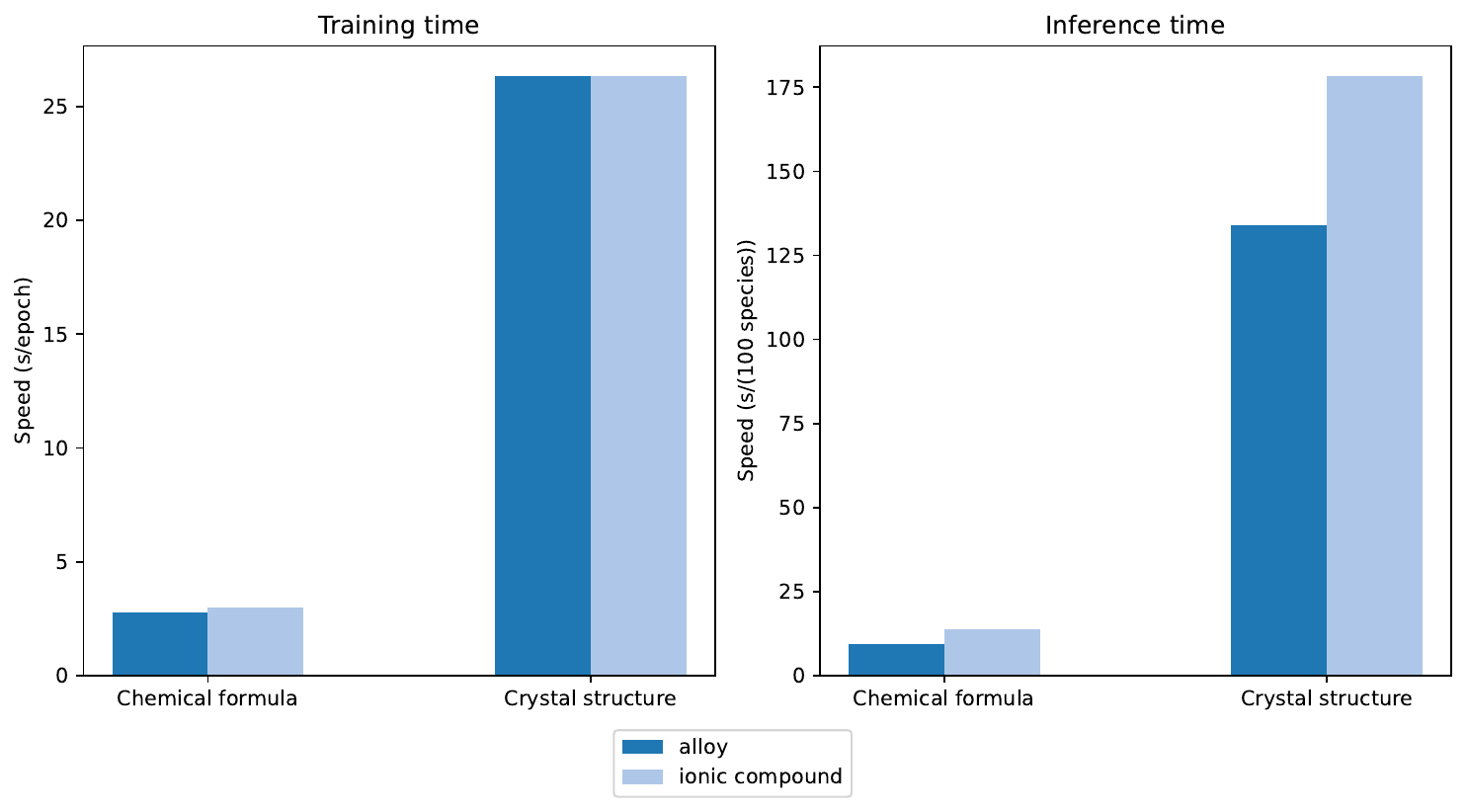}
    \caption{Comparison between the speed of chemical formula and crystal structure generation.}
    \label{fig:speed}
\end{figure}

In Fig.\,\ref{fig:speed}, we test the speed of chemical formula generation with the chemical formula transformer and the crystal structure generation with the crystal diffusion model. For the training stage, models are trained on the training partition of the MP-20-valence dataset. Because a unified crystal diffusion model is trained for both alloys and ionic compounds, they share the same training speed in the figure. For the inference stage, the chemical formula transformer samples chemical formulas one by one until 100 distinct species are obtained. They are then fed into the unconditioned crystal transformer model to generate one structure for each species. The results indicate that the chemical formula generation as an additional prefix of the material generation task in our framework doesn't significantly increase the computation time.

\subsection{Validation of energy above hull using MLIP}

To assess the reliability of the MatterSim MLIP in evaluating the quality of generated crystals, we first benchmark its predictions on known structures from the MP-20 training dataset. Since all entries in this dataset are known to have energy above hull values below 0.1 eV (as computed via DFT), this serves as a consistency check for the MLIP. As shown in Fig.\,\ref{fig:ehull}, the MatterSim-evaluated $E_{hull}$ values cluster tightly below the 0.1 eV threshold, demonstrating strong agreement with DFT results. This validates MatterSim’s ability to serve as a proxy for rapid stability assessment and supports its use in evaluating newly generated crystals.

\begin{figure}[!htbp]
    \centering
    \includegraphics[width=0.6\linewidth]{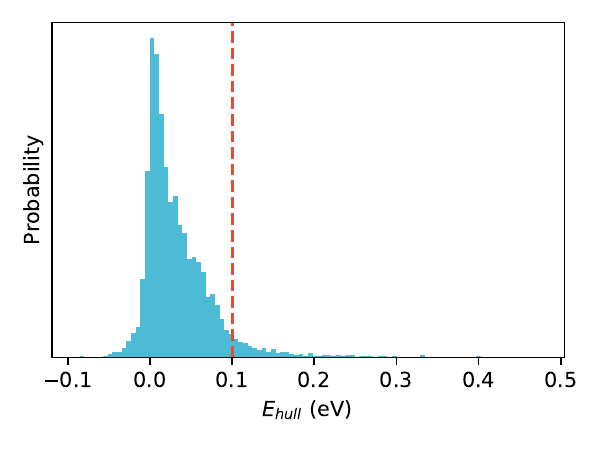}
    \caption{Distribution of energy above hull for crystals in MP-20 dataset evaluated using machine learning interatomic potential. MatterSim is used for on-the-fly evaluation, with the red dashed line marking the 0.1 eV threshold commonly used for assessing thermodynamic stability.}
    \label{fig:ehull}
\end{figure}

\subsection{Validation of thermal conductivity using MLIP}
In this section, we benchmark the fast thermal conductivity evaluation in our CrysVCD inverse design workflow using a MatterSim MLIP model. While MatterSim MLIP has been validated on the Matbench-Discovery benchmark \cite{riebesell2023matbench}, the present test targets a biased subset: candidate materials with human-guided high thermal conductivity, typically containing light elements. 
This bias naturally shifts the distribution towards higher $\kappa$, where errors could tend to grow. For this dataset, the symmetric relative mean error ($\kappa_{RMSE}$) in total thermal conductivity predictions is 0.86, larger than the value (0.58) reported in MatterSim for general materials. Nevertheless, the model retains a reasonable correlation with DFT reference values and demonstrates practical reliability for rapid screening in high-$\kappa$ design scenarios.

\begin{figure}[!htbp]
    \centering
    \includegraphics[width=0.4\linewidth]{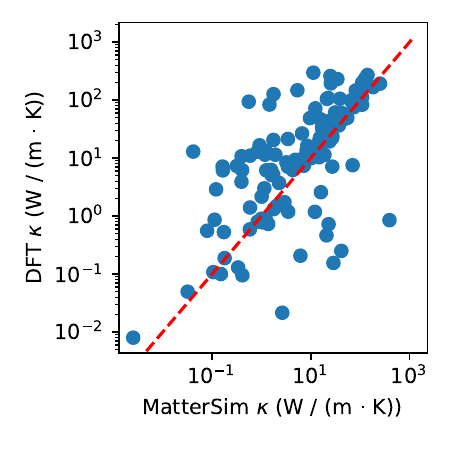}
    \caption{Comparison between the thermal conductivity $\kappa$ computed by DFT and MatterSim MLIP. The difference lies in the supplement of force constants, either calculated by DFT or evaluated by MLIP, and both results are subsequently passed on to phono3py for thermal transport calculation.}
    \label{fig:ehull}
\end{figure}

\subsection{Surrogate GNNOpt model for dielectric constant prediction}
\begin{figure}[!htbp]
    \centering
    \includegraphics[width=0.7\linewidth]{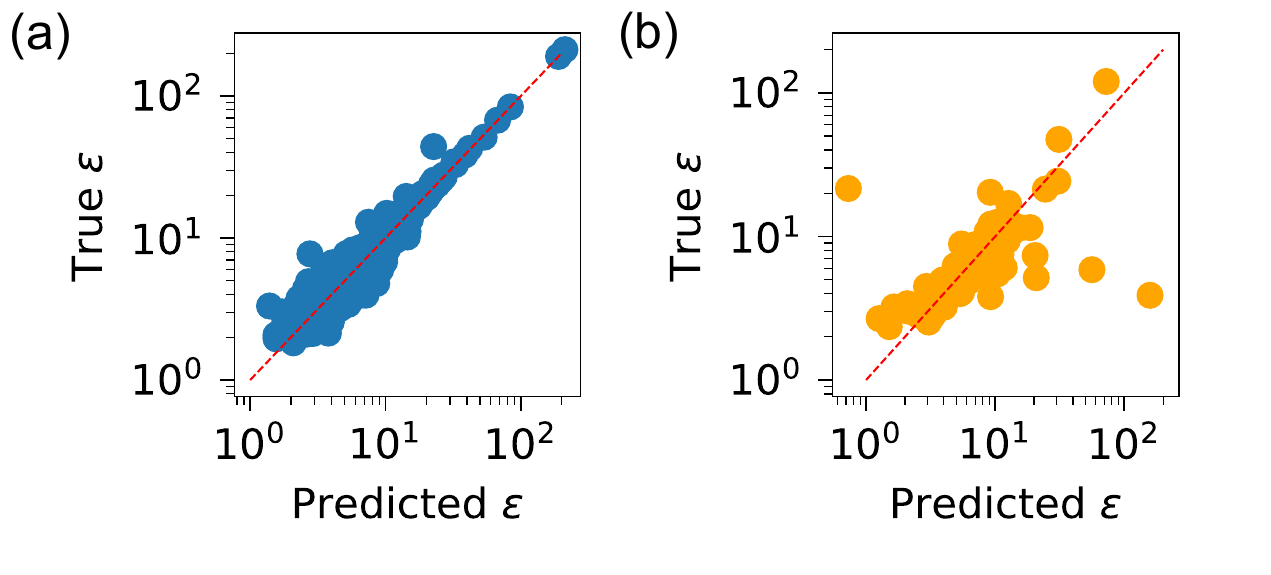}
    \caption{Predicted vs. true static dielectric constant ($\varepsilon$) using an E(3)-equivariant graph neural network (e3nn) model. \textbf{(a)} Training set and \textbf{(b)} testing set results are shown with a red dashed line representing the ideal parity line.}
    \label{fig:opt}
\end{figure}

To enable efficient inverse design of high-$\kappa$ dielectric materials, we employ an E(3)-equivariant graph neural network (e3nn) as a surrogate model to predict the static dielectric constant $\varepsilon$, based on the implementation of Hung et al.\cite{hung2024universal}. 
Given the high computational cost of direct DFT evaluation, this model allows for rapid screening of generated candidate materials. As shown in Fig.\,\ref{fig:opt}, the model demonstrates good agreement with ground-truth values on the training set and maintains reasonable accuracy on the testing set. The predictive performance justifies the suitability of the e3nn model for evaluating dielectric properties within our generative design framework.

\newpage
\bibliography{refs_intro.bib}